# Sonication-assisted liquid phase exfoliation of two-dimensional CrTe$_3$ under inert conditions

Kevin Synnatschke,[a,b] Narine Moses Badlyan,[c,d] Angelika Wrzesińska,[e,f] Guillermo Lozano Onrubia,[a] Anna-Lena Hansen,[g,I] Stefan Wolff,[d] Hans Tornatzky,[c,h] Wolfgang Bensch,[I] Yana Vaynzof,[e,f] Janina Maultzsch,[c,d] Claudia Backes[j*]

[a] Institute of Physical Chemistry, Heidelberg University, Im Neuenheimer Feld 253, 69120 Heidelberg, Germany
[b] School of Physics, University of Dublin, Trinity College, Dublin 2, Ireland
[c] Institute for Solid-State Physics, Technische Universität Berlin, Hardenbergstraße 36, 10623 Berlin, Germany
[d] Department of Physics, Friedrich-Alexander-Universität Erlangen-Nürnberg, Staudtstr. 7, 91058 Erlangen, Germany
[e] Chair for Emerging Electronic Technologies, TU Dresden, Nöthnitzer Str. 61, Dresden, 01187 Sachsen, Germany
[f] Leibniz-Institute for Solid State and Materials Research Dresden, Helmholtzstraße 20, Dresden 01069, Sachsen, Germany
[g] Institute for Applied Materials—Energy Storage Systems (IAM-ESS), Karlsruhe Institute of Technology (KIT), 76344 Eggenstein, Germany
[h] Paul-Drude-Institut für Festkörperelektronik, Leibniz-Institut im Forschungsverbund Berlin e.V, Hausvogteiplatz 5–7, 10117 Berlin, Germany
[i] Institute of Inorganic Chemistry, University of Kiel, Max-Eyth-Straße 2, 24118 Kiel, Germany
[j] Chair of Physical Chemistry of Nanomaterials, University of Kassel, Heinrich-Plett-Straße 40, 34132 Kassel, Germany

Email: backes@uni-kassel.de

Liquid phase exfoliation (LPE) has been used for the successful fabrication of nanosheetsfrom a large number of van der Waals materials. While this allows to study fundamental changes of material properties' associated with reduced dimensions, it also changes the chemistry of many materials due to a significant increase of the effective surface area, often accompanied with enhanced reactivity and accelerated oxidation. To prevent material decomposition, LPE and processing in inert atmosphere have been developed, which enables the preparation of pristine nanomaterials, and to systematically study compositional changes over time for different storage conditions. Here, we demonstrate the inert exfoliation of the oxidation-sensitive van der Waals crystal, CrTe$_3$. The pristine nanomaterial was purified and size-selected by centrifugation, nanosheet dimensions in the fractions quantified by atomic force microscopy and studied by Raman, X-ray photoelectron spectroscopy (XPS), energy-dispersive X-ray spectroscopy (EDX) and photo spectroscopic measurements. We find a dependence of the relative intensities of the CrTe$_3$ Raman modes on the propagation direction of the incident light, which prevents a correlation of the Raman spectral profile to the nanosheet dimensions. XPS and EDX reveal that the contribution of surface oxides to the spectra is reduced after exfoliation compared to the bulk material. Further, the decomposition mechanism of the nanosheetswas studied by time-dependent extinction measurements after water titrationexperiments to initially dry solvents, which suggest that water plays a significant role in the material decomposition.

1. **Introduction**

In the past decades, two-dimensional (2D) materials gained increasing attention due to unique properties and novel physics arising from quantum confinement effects. While theoretic predictions help to identify interesting compounds for a range of applications, 2D magnetic materials have been studied intensively in recent years [1], [2], [3], [4], [5], [6], [7], [8], [9], [10], [11], [12], [13], [14]. However, both, the synthesis and exfoliation of such material systems can sometimes be challenging, which can partially be attributed to extreme conditions for the formation of distinct polymorphs, or their sensitivity to environmental conditions [1], [8], [15], [16]. A layered antiferromagnet, $CrTe_3$ was first reported in 1979 [17]. While theoretic considerations predict the material to be interesting for studies on the magnetic ordering in low-dimensional structures [3], [18], experimental data is scarce [3], [19], [20] due to challenges associated with phase puresynthesis, control of the dimensions and poor material stability [21], [22]. The preparation of phase-pure $CrTe_3$ is a challenge because both the Cr:Te ratio and the heating procedure must be precisely controlled. The information concerning the synthesis conditions of $CrTe_3$ is somehow conflicting, but for an exactly stoichiometric sample, peritectic decomposition occurs at 753 K.

To study magnetism in 2D materials, samples are typically produced by micromechanical exfoliation of single crystals or bottom up growth by chemical vapour deposition or molecular beam epitaxy [14], [19], [20]. Recently, using the sulfosalt cylindrite (approximate composition $Pb_3Sn_4FeSb_2S_{14}$) [23] or $Fe_2P_2S_6$ [8], [24] as model substances, it was suggested that liquid-suspended nanosheets might also constitute a suitable sample type to assess magnetic ordering. A harsh, yet widely applied approach for the production of nanosheets in the liquid environment is liquid phase exfoliation [25], [26], [27]. Here, layer separation is achieved through sonication or shearing and the resultant nanomaterial is stabilized against aggregation through appropriate solvents with matching solubility parameters or surfactants as stabilizers adsorbed on the surface [25], [26], [27]. In recent years, this method has become increasingly popular due to its applicability to a broad range of material classes including non-layered materials [28], [29], [30]. Importantly, no large area single crystals are required for LPE (in contrast to micromechanical exfoliation) and bulk materials with a powder-like texture and small grain sizes are equally suitable [31]. Thus LPE might indeed be an alternative for the production of mono- or few-layered nanosheets of $CrTe_3$ that are not readily accessible by other means.

LPE is a relatively crude process and as such produces nanosheets with broad size and thickness distributions [32], [33]. To address this, various size selection techniques have been developed. A popular approach are centrifugation-based methods [34], such as density gradient centrifugation (DGU) [35], [36], [37] or liquid cascade centrifugation (LCC) [21], [27], [33], [38]. In short, DGU gives a more precise control over the thickness of the isolated nanomaterial, but lacks scalability and requires additives to tune the density of the centrifugation medium [36], [39] which may impact the colloidal stability [40], [41], [42], and are often difficult to remove. LCC gives poorer control over the size and thickness of the nanosheets, but bears the advantage that larger amounts of material can be handled without any additional components being added to the system [27]. In addition, LCC has been successfully applied to dispersions of different nanomaterials in different liquid environments including solvents or aqueous surfactant systems which has enabled the systematic study of size and thickness dependent properties [15], [21], [27], [43], [44], [45]. Furthermore, it was shown that the size selection by LCC removes oxides and other impurities as soluble decomposition products, as well as highly defective nanomaterial in the supernatant after centrifugation [15], [27], [46].

In spite of such size selection techniques, the lateral dimensions of nanosheets from LPE are typically limited. This can be rationalised by in-plane nanosheet tearing which occurs during the high energy treatment in LPE in addition to the desired out-of-plane delamination [33], [47], [48]. While it was suggested that the nanosheet length and thickness aspect ratio is governed by the in-plane to out-of-plane binding strength [33], [48] and thus in principle accessible through theoretical predictions, the accuracy of calculations may be limited in the case of complex structures and magnetic materials. It is therefore important to experimentally assess the ease of exfoliation for currently little explored materials such as $CrTe_3$.

Furthermore, the scission of covalent bonds and creation of new edges can result in an oxidation of the material, even in the case of graphene [47], [49], [50] which is relatively inert towards oxidation in ambient conditions. This can be even more crucial for other layered materials that are more prone to oxidation such as transition metal dichalcogenides [45], [51], [52], [53] or black phosphorus [43], [54], [55]. For any new material exfoliated *via* LPE, it is thus important to investigate the chemical stability over time regardless of the intended application area. Since the material's reactivity may additionally vary with nanosheet size and thickness due to changes of the effective material surface area, but also due to a different chemical behaviour of material edges and basal planes [51], [56], it is beneficial to perform such studies as function of nanosheet size and thickness. As previously shown for black phosphorus [43], TiS$_2$ [45], MoO$_2$ [44], Ni$_2$P$_2$S$_6$ [15], and titanium carbide [57], the degradation of LPE nanosheets can be readily followed by measuring absorbance or extinction spectra of the colloids as function of time. This constitutes a relatively simple approach to assess the chemical stability of the nanosheets, for example in ambient conditions and allows to identify timeframes for further advanced characterisation or processing.

Here we show that sonication-assisted LPE and LCC-based size selection can be applied to bulk CrTe$_3$ in *N*-cyclohexyl-2-pyrrolidone as solvent under inert gas conditions. The composition is analysed through XPS and EDX. The nanosheet dimensions in the size-selected fractions are quantified through statistical atomic force microscopy. Size-dependent optical properties are evaluated by Raman spectroscopy, as well as optical extinction spectroscopy. Finally, nanosheet degradation is followed through extinction spectroscopy as function of time after exposure of the samples to water and oxygen.

## 2. Results and Discussion

### 2.1. CrTe$_3$ bulk Characterisation

CrTe$_3$ crystals were prepared by solid state diffusion methods at high temperatures as reported previously [22]. A schematic crystal structure of the compound is shown in Fig. 1A [22], [58]. The dashed lines mark one Cr$_4$Te$_{16}$ unit which is the smallest repeating unit across the 2D-plane. The structure consists of four edge-linked CrTe$_6$ octahedra, connected to four neighbouring groups by apical Te-atoms. This way, the Cr$_4$Te$_{16}$ units arrange in a zig-zag fashion along the layered plane, as illustrated in the structure on the right in Fig. 1A [22].

Powder diffraction measurements (PXRD) show no other than the expected Bragg reflections, reported in literature (Fig. 1B) [16], [22]. The diffraction data indicates that the starting material is phase-pure and clean from crystalline impurities. The bulk material shows the characteristic layered structure, albeit with relatively small crystals of a typical size between 2 and 10 µm as determined by scanning electron microscopy (SEM, Fig. 1C-D). A zoom-in to a single crystal reveals the individual layers (Fig. 1C–D). For further analysis, X-ray photoelectron spectroscopy (XPS) and Raman measurements on the bulk material were performed. The XPS Te3d and core level spectra, also including the overlap with the Cr2p core level, (Fig. 1E) reveal evidence of surface oxidation from fitting the Te3d signals. The oxide contribution from the core level fit is estimated to be approximately 38%, indicating a significant level of surface oxidation. Due to the overlap of the Te3d signals with the Cr2p core level, this is to be considered as an estimate due to uncertainties in deconvoluting the peaks. We note that, while the bulk material was stored and shipped under $N_2$ atmosphere, some exposure to the environment occurred on transfer to the XPS chamber. In spite of this surface oxidation, we emphasize that the crystals predominantly contain phase-pure $CrTe_3$ according to the XRD.

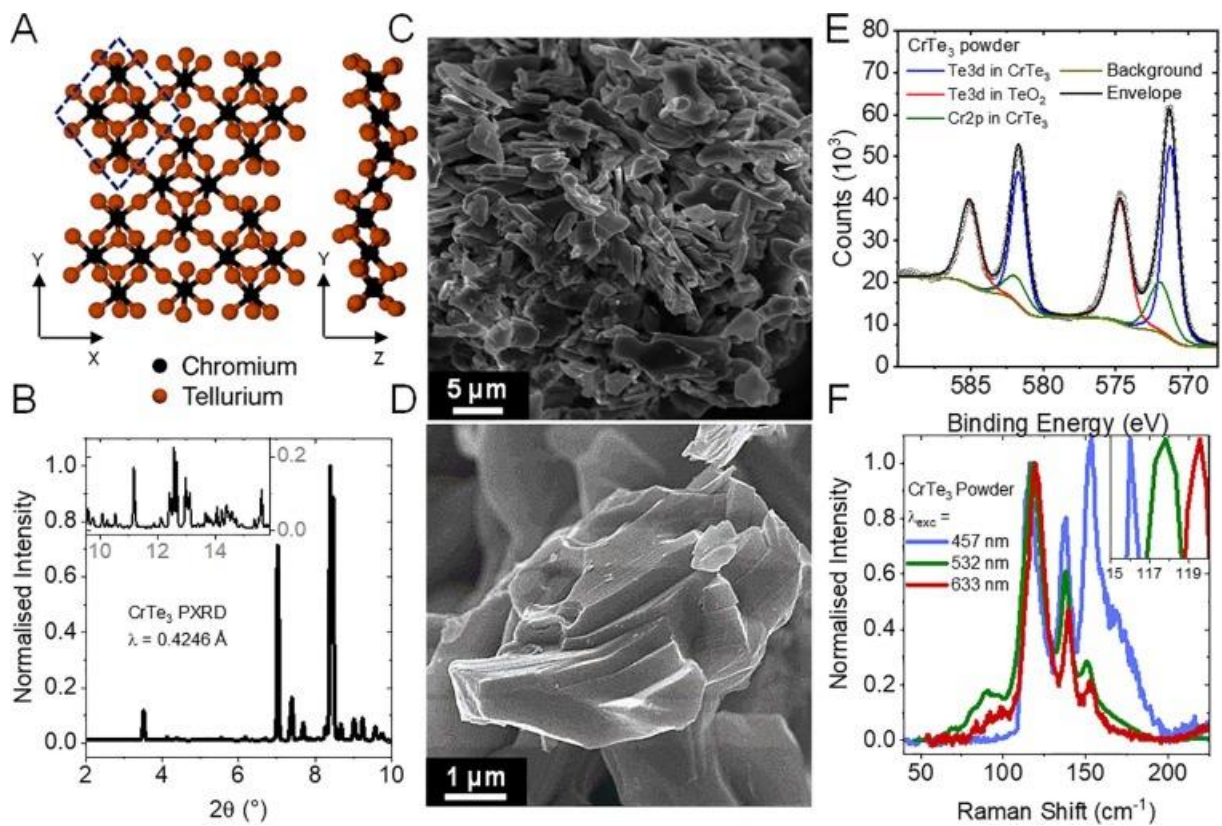

**Fig. 1.** Characterisation of bulk $CrTe_3$. (**A**) Schematic crystal structure. Individual layers (x/y–plane) are made up of edge sharing $Cr_4Te_{16}$ clusters (dashed rhombus), stacked along the z-axis. (**B**) Powder diffraction of bulk $CrTe_3$ (T = 294 K). All reflections can be assigned to $CrTe_3$ [22]. (**C**) Low magnification SEM image of the starting material showing crystallites of few microns (∼2–10 µm) in size. (**D**) Higher resolution SEM images of a crystal (**C**). The layered character of the material is resolved. (**E**) Fitted Te3d and Cr2p core level spectra of the starting material. Note that oxide species are identified in the starting material suggesting that 38% of the tellurium atoms stem from $TeO_2$ rather than $CrTe_3$. (**F**) Raman spectra of bulk $CrTe_3$ for excitation with a 457, 532 and 633 nm laser normalised to the vibrational mode at ∼ 118 $cm^{-1}$ for better comparison. Slight shifts of the peak position for three different excitation wavelengths are observed (inset) in addition to changes in relative peak intensities.

In addition, Raman measurements were carried out on the starting material using different excitation wavelengths (457, 532 and 633 nm). The spectra show three well-defined, relatively broad signals for all excitation wavelengths. (Fig. 1F). Additionally, a systematic red-shift of the mode at 115–120 cm$^{-1}$ is observed with increasing excitation energy (see inset, Fig. 1F). Note that the Raman spectra of CrTe$_3$ have not been discussed in any more detail in literature to date. Hence, the modes cannot be assigned without additional theoretical efforts. This will be addressed further below.

*2.2. Liquid-Phase exfoliation and size selection*

To test the exfoliability of CrTe$_3$, LPE was first performed by tip sonication, using aqueous sodium cholate solution. However, strong aggregation and material decomposition was noticed in preliminary atomic force microscopy (AFM) and in optical extinction measurements. Thus, the exfoliation was performed in inert gas conditions in organic solvent. For this purpose, ground crystals were bath-sonicated in an argon atmosphere, using distilled, dried, and degassed *N*–cyclohexyl-2-pyrrolidone (CHP, see methods section and SI for experimental details), as a medium. Pyrrolidone-based solvents are known to form stable dispersions of various nanomaterials due to matching solubility parameters which can be described in the framework of solution thermodynamics [59]. In addition, these solvents can efficiently protect the nanosheets from environmental degradation due to strong interactions between the solvent and the nanomaterial surface as shown for other materials prone to oxidation, such as TiS$_2$ and black phosphorus [43], [45].

To assess the structural integrity of the exfoliated nanomaterial, scanning electron microscopy (SEM) was performed on a deposited nanosheet dispersion (without size selection, termed "stock-like"). For this purpose, unexfoliated material was removed by centrifugation at low centrifugal accelerations (100 *g* where *g* denotes multiples of the earth's gravitational field, 2 h). A second centrifugation at higher centrifugal accelerations (30k *g*, 2 h) was performed on the exfoliated material decanted as supernatant from the previous step. After this centrifugation at higher accelerations, the sediment was redispersed and collected for analysis This step was performed to i) allow for redispersion at reduced volume and thus higher nanosheet concentration and ii) to remove impurities from solvent degradation and very small, defective nanosheets as described for Ni$_2$P$_2$S$_6$ and α-RuCl$_3$ previously [15], [60]. The exfoliation yield for such a CrTe$_3$ sample was determined gravimetrically (see Methods) as ∼ 10%.

The concentrated nanomaterial was deposited on Si wafers for SEM and electron dispersive X-ray (EDX) measurements, as well as ITO substrates for XPS measurements, respectively by drop casting in a nitrogen atmosphere. The SEM measurements reveal a broad distribution of nanosheets with varying sizes and thicknesses (inset Fig. 2A; wide view images shown in the SI, Fig. S2A-B). The analysis of the EDX spectra of exfoliated nanosheets in comparison to the bulk material is shown in Fig. 2A. In spite of an overlap of the Cr$_{Lα}$ and O$_{Kα}$ lines, the data suggests a lower oxygen content in the nanosheets compared to the starting material, with an EDX-suggested Cr:Te:O ratio of 1:3.1:0.6 for the nanomaterial, while the ratio for the starting material is approximated as 1:2.7:3.2. This implies that no additional oxidation occurred on exfoliation, but rather that the oxide content is reduced due to the centrifugation-based purification as previously observed for other materials [15], [60].

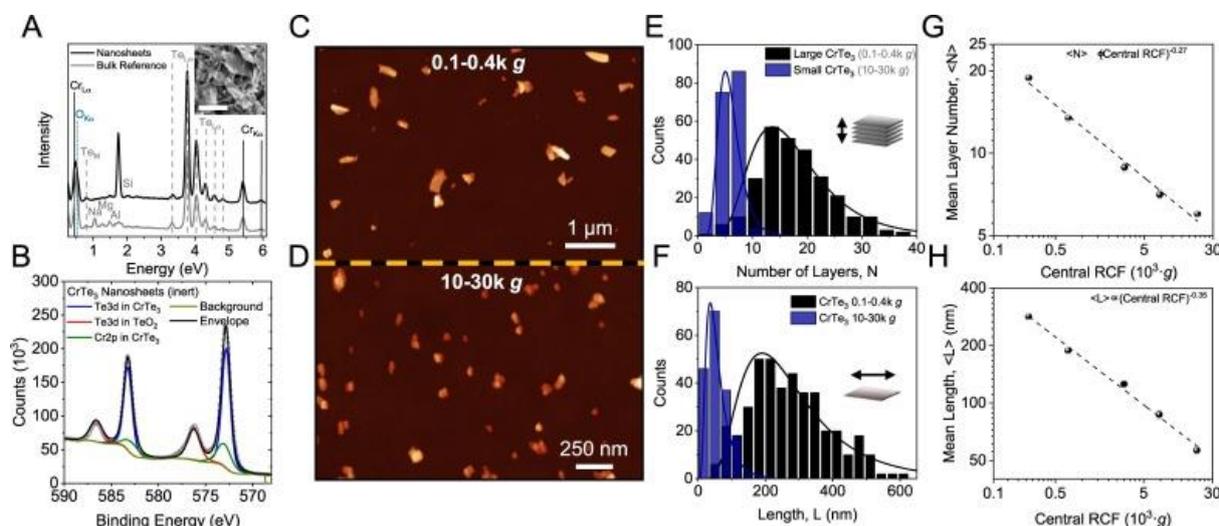

**Fig. 2.** Characterisation of exfoliated CrTe$_3$ nanosheets. **(A)** EDX spectrum of exfoliated nanosheets (black) in comparison to the starting material (grey). All peaks in the spectrum for the nanomaterial can be attributed to CrTe$_3$ and impurities from Na, Mg and Al the starting material. The inset shows an SEM image of few deposited nanosheets. A larger overview is shown in the SI, Figure S2A–B. The scalebar is 400 nm. **(B)** Fitted Te3d and Cr2p core level spectra of the exfoliated nanomaterial. Note that both, EDX, and XPS measurement suggest that the oxide contribution in the starting material is higher than in the exfoliated material. **(C-D)** AFM images of drop-cast nanosheets for two fractions containing relatively large nanosheets, RCF = 0.1–0.4k $g$ **(C)** and small sheets, RCF = 10–30k $g$ (D). **(E-F)** histograms of the nanosheet layer number (**E**) and the lateral sheet size (**F**) distribution for the two fractions shown in **C** + **D**. For other fractions see SI, Figure S3). **(G)** Arithmetic mean layer number, as function of central RCF on a double logarithmic scale. **(H)** Arithmetic mean lateral size, as function of central RCF on a double logarithmic scale. To calculate , the AFM apparent height is converted into layer number by step height analysis on terraces of sheets (See SI, Figure S2C–D for details). The dashed lines are power law fits for both, and , respectively as a function of the central centrifugal acceleration.

To confirm this, XPS was performed for the exfoliated material. Fitting of the Te3d and Cr2p core level spectra reveals a decrease of the oxygen contribution to the Te3d peaks from 38% to 23.5% when comparing the starting material to the exfoliated species (Fig. 2B). Results from both, EDX and XPS measurements thus demonstrate that the above-anticipated purification through centrifugation was successful leading to a reduction of the content of surface oxides.

Due to the typically polydisperse nature of LPE samples in terms of nanosheet dimensions, microscopy statistics are required to assess the exfoliation quality, *i.e.* length/thickness aspect ratio. AFM has proven to be a suitable technique, as lateral dimensions and the material thickness can be determined simultaneously [61], [32], [33]. However, it is important to adjust the scanned area to the size of the nanosheets in order to apply corrections for pixilation and cantilever broadening for an accurate size determination [21]. To achieve this, it is beneficial to perform a size selection to narrow the distribution width of the dimensions. To this end, LCC was applied to the stock dispersion of exfoliated $CrTe_3$ in CHP after sonication (see methods for details). The dispersion was subjected to consecutive centrifugation steps at increasing centrifugal accelerations in analogy to previous work [27]. After each step, supernatant and sediment were separated, the supernatant subjected to the subsequent centrifugation step at higher acceleration and the respective sediment redispersed in fresh solvent and subjected to analysis. We note that the sediment was collected in reduced solvent volume to facilitate further analysis. Since the first sediment and the last supernatant typically contain predominantly unexfoliated and defective material, (see above) these two fractions were discarded. Isolated fractions of different size are labelled by the centrifugal acceleration applied between two consecutive centrifugation steps. The respective steps for the size selection of $CrTe_3$ were 100 $g$, 400 $g$, 1000 $g$, 5000 $g$, 10000 $g$ and 30000 $g$. A schematic illustration of the procedure is shown in the supporting information (Figure S1).

The concentrated nanosheet fractions in CHP obtained after LCC were diluted in nitrogen atmosphere, using dry, degassed, and distilled isopropanol, and immediately deposited on $Si/SiO_2$ wafers (300 nm oxide layer) by flash evaporation for the AFM measurements (see experimental section for more details). Representative nanosheets of the largest (0.1–0.4k $g$) and the smallest isolated size fraction (10–30k $g$) are shown in Fig. 2C–D. For each fraction, >300 nanosheets were analysed and the longest dimension (length, L), dimension perpendicular (width, w) and thickness measured manually using line profiles over the sheets. The nanosheet lateral dimensions were corrected for cantilever broadening and pixilation effects using previously established empirical calibrations [21]. In addition to cantilever broadening, the thickness of LPE nanosheets is typically overestimated due to contributions from solvent residues, an impact from measurement parameters (set point, free amplitude, *etc.*) and nanosheet surface properties [62], [63]. Thus, the measured height, is proportional to, but not identical to the product of layer number and crystallographic thickness of one layer. However, the apparent height can be converted into layer number of the 2D material by step height analysis as demonstrated for different materials [21], [43], [56], [64]. The procedure involves the measurement of the AFM height of steps associated with terraces of incompletely exfoliated nanosheets (see Figure S2C for an example). Sorting the step heights in ascending order reveals discrete steps, in this case of 1.9 nm (Figure S2D), which can be assigned to the

AFM thickness of a single layer of CrTe$_3$ exfoliated in CHP. Hence, for conversion of the apparent height (T) to layer number (N), the measured thickness is divided by 1.9 nm. This statistical analysis allows to construct histograms of the corresponding nanosheet layer number and lateral size distribution as exemplarily shown for the smallest and largest fraction in Fig. 2E–F (for all fractions see supporting information, Figure S3).

From the distributions of each fraction, the arithmetic mean layer number (<N>) and length (<L>) was calculated. To visualise the size selection, these arithmetic averages are plotted as function of the midpoint of the pair of centrifugal acceleration used to isolate the respective fraction on a double logarithmic scale in Fig. 2G–H. As expected for this centrifugation based size selection [27], [65], a typical power law dependence is observed as indicated by the dashed lines. For CrTe$_3$ investigated here, fitting yields a quantitative relation of the average dimensions as  and . We note that the observed scaling is not quantitatively consistent with the expected −0.5 exponent according to centrifugation theory initially developed for isotopically shaped particles [66], even though it was shown to hold for other 2D materials, such as graphite and transition metal dichalcogenides (TMDs) [27].

This suggests additional influences on the material sedimentation velocity beyond the current understanding. An experimental error (*e.g.*, non-uniform decanting inside the glovebox) cannot be excluded even though this is not expected to have an impact on the scaling of the exponents with *RCF*. Nonetheless, it is possible that defined exponents for material systems handled in inert gas conditions might be difficult to compare due to additional processing steps, which are a potential source of experimental errors (*e.g.*, shaking of centrifuged samples when passing through the glovebox antechamber). However, we note that CrTe$_3$ is not an exception, as similar deviations from the expected scaling have been observed for other material systems, such as *h*-BN [67], and Ni$_2$P$_2$S$_6$[15]. A potential reason could be a degradation-induced change of the solvation shell and thus the buoyant density. If this process occurred during the centrifugation, the exponent would also be affected. In addition, aggregation during centrifugation would also affect the scaling exponent. Overall, the exact origin of these variations in the exponent across different material systems remains elusive and additional data on different material systems and solvents will be required in combination with improved models for the sedimentation behaviour of platelets.

In contrast to fraction-averaged material dimensions which are dependent on the sedimentation behaviour, the overall population of sizes and thicknesses measured in all material fractions can be regarded as representative for the size and thickness range of CrTe$_3$ nanosheets produced by sonication-assisted LPE. This data is summarized in Figure S4. Since the relationship of lateral size and thickness can be considered as a measure for the material exfoliability [33], the area of all individual nanosheets measured by AFM is shown as function of their thickness in a scatterplot (Figure S4 B). Clearly, larger nanosheets tend to be thicker and smaller nanosheets tend to be thinner. It was previously suggested that the scaling of area with thickness can be used to extract a material-dependent parameter termed characteristic monolayer size which allows for a numerical comparison of the ease of exfoliation with other materials [33]. To obtain this value, nanosheets were grouped by thickness and the average area calculated. The data is shown in Figure S4C and displays a characteristic power law scaling which follows . Note, that the exponent is numerically different from fraction-averaged data which was previously used in literature [33]. Thus, for a better comparability, previously reported data of graphene, $WS_2$ and $MoS_2$ nanosheets was re-evaluated in a similar way which leads to the same exponent as observed for $CrTe_3$ (Figure S4 C). Extrapolation of the fit to layer number 1 allows to determine the characteristic monolayer size of $CrTe_3$, = 28.4 nm, which is significantly smaller than the characteristic monolayer size of graphene (97.2 nm), $WS_2$ (39.8 nm) and $MoS_2$ (41.5 nm). This clearly shows that thin $CrTe_3$ nanosheets are more difficult to produce by LPE compared to widely studied systems.

*2.3. Size dependent material properties*

Raman experiments were carried out on different sizes of exfoliated nanosheets to obtain deeper insights into size-dependent material properties. The material was drop-cast into a conic indentation on a gold-coated aluminium substrate and encapsulated in a nitrogen atmosphere before the measurements for this purpose. As observed for the bulk material, three distinct peaks are found for all sizes of the exfoliated nanosheets (Fig. 3A-B). However, differences between measurements acquired at different positions of the same sample were observed as exemplarily shown in Fig. 3A-B for measurements of comparatively small (5–10k *g*, A) and large (0.1–0.4k *g*, B) nanosheets at two different positions, respectively. Slight shifts, as well as differences in intensity ratios can be observed for measurements at the centre, compared to measurements at the edge of the drill hole where the nanosheets were deposited in. The effect is more pronounced for the 5–10k g fraction. A possible rationale could be a dependence on the Raman spectrum on the alignment of the platelets relative to the incident beam.

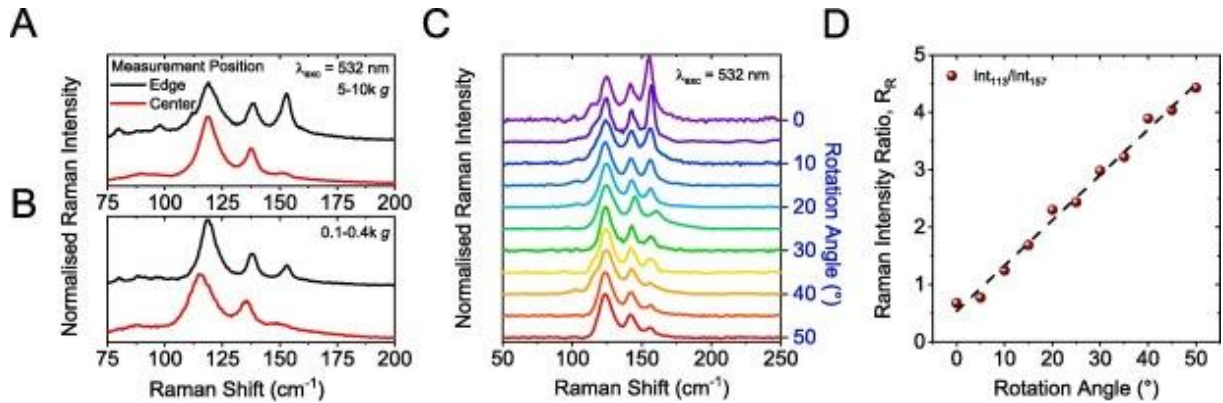

**Fig. 3.** Tilt-angle dependence of CrTe$_3$ phonon modes. **(A-B)** Raman measurements (excitation wavelength 532 nm) for different sizes of LPE nanosheets acquired at different positions of the substrate (*i.e.,* drill hole, see Figure S5). Both, relatively small **(A)** and relatively large **(B)** nanosheets show distinct signals for measurements at the edges and for measurements at the centre of the hole. **(C)** Raman spectra of a CrTe$_3$ crystal for different tilting angles. **(D)** Intensity ratio for the phonon modes at 113 and 157 cm$^{-1}$. The linear trend is strictly empirical and is not correlated to crystallographic data, but serves the purpose to demonstrate the dependence of the phonon modes with the tilt angle of the incident light for excitation.

It is therefore crucial to investigate the angle dependence between the incident light beam used for the excitation of vibrational modes and the CrTe$_3$ crystals. To achieve this, a custom-built sample holder was used and the surface of a comparatively large bulk crystallite was cleaned by repeatedly applying and removing adhesive tape to the crystal surface. The same position of the cleaned sample surface was repeatedly measured and tilted, applying 5° steps after each measurement in a total range of 50° (Fig. 3C). A systematic change of the Raman intensity ratios at ~ 115 cm$^{-1}$ and ~ 155 cm$^{-1}$ with the direction of incident light is observed with progressing sample rotation (Fig. 3D). A roughly linear scaling of the intensity ratio within the studied tilting range is observed. We note that the experimental setup used for the sample tilting is prone to experimental errors and the observed trends might not be fully quantitative. Unfortunately, no information can be provided about the crystallographic orientation of the crystallite. Absolute changes of the sample's tilt angle are thus not accessible, but the measurements show that the spatial orientation of the crystal has a strong impact on the observed phonon excitation. The results obtained for the bulk material are in good agreement with the spectral changes observed for different positions on the drop-cast nanomaterial in the indent. A direct comparison of the peak intensity ratios of the two different positions measured for the small exfoliated nanosheets (Fig. 3A) to the results presented in Fig. 3D, suggests tilting angles of about 10 and 30° for the exfoliated small nanosheets, respectively which fits roughly to the geometry of the conical indent, in which the material was deposited (see Figure S5). This is an interesting observation, as it implies that the nanosheets align on the substrate surface even for confined deposition from dispersion at high concentrations. In addition, it emphasizes that care needs to be taken to evaluate spectroscopic changes across samples in the case of nanosheet ensembles produced from LPE.

To understand the Raman data in more detail, a density functional perturbation theory(DFPT) approach was applied on both, monolayered and bulk $CrTe_3$ to calculate phonon resonances. The results are compared to experimental results on the bulk powder and a fraction containing comparatively small and thin nanomaterial, respectively (Fig. 4A-B). While some differences between measured and calculated modes are observed, shifts of transition frequencies and changes of the absolute intensity are confirmed by the theoretical approach.

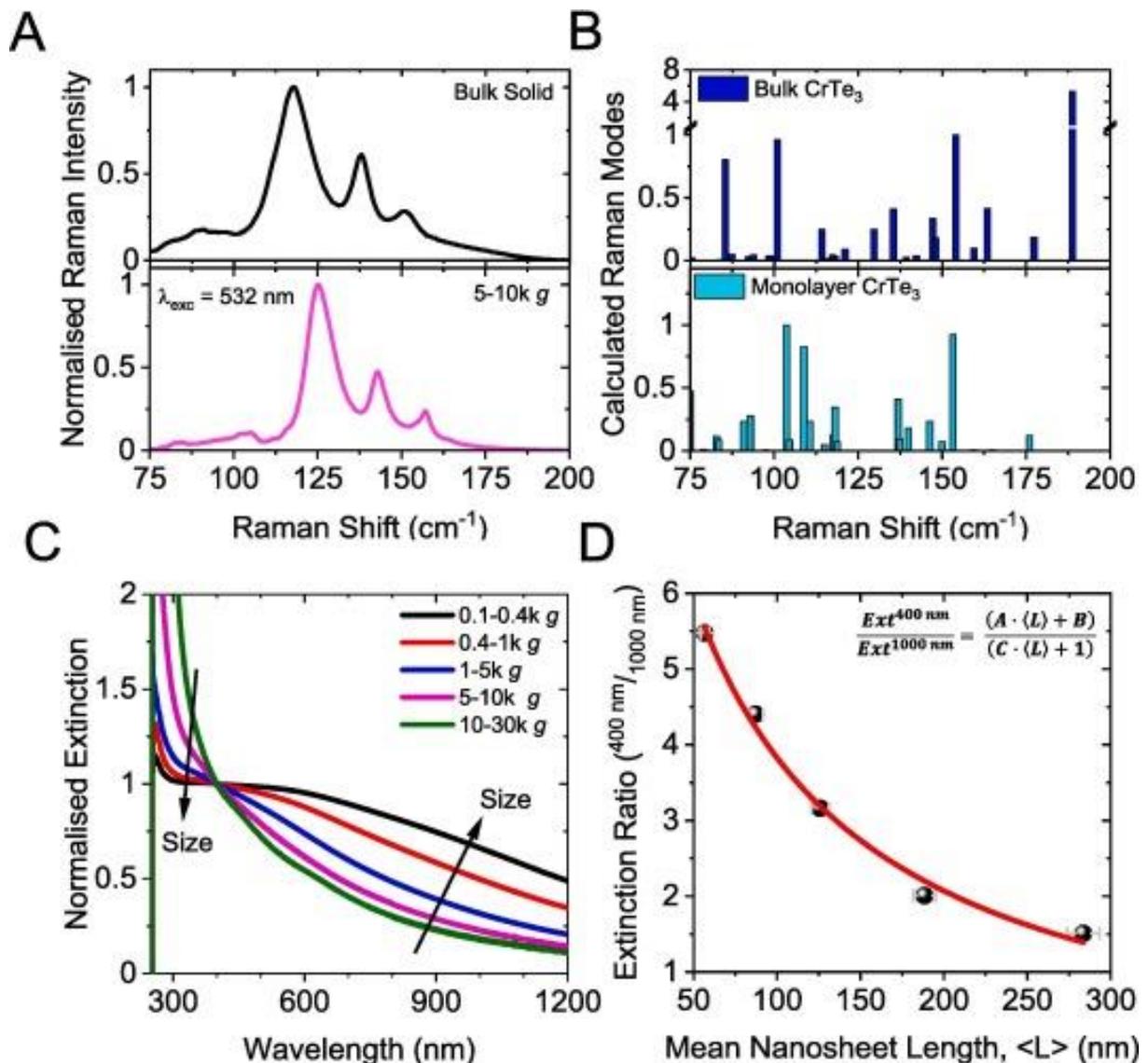

**Fig. 4.** Size-dependent optical response of $CrTe_3$. **(A-B)** Raman measurements (**A**) and DFT calculations (**B**) for bulk solid (top) and delaminated (bottom) $CrTe_3$. Both, theory and experiment show changes of the phonon modes with material thickness. **(C)** Extinction coefficient spectra of $CrTe_3$ dispersions with varying size- and thickness distributions from the ultraviolet to nIR range of the electromagnetic spectrum. A systematic change is observed for the extinction coefficient with changing material dimensions. **(D)** Intensity ratio at 400 nm

over 1000 nm serves as metric for the lateral size and is plotted as a function of <L>. The red line is a fit to equation 3.

In addition to Raman measurements, different sizes of CrTe$_3$ nanosheets were subjected to extinction measurements (Fig. 4C). Systematic changes of the extinction spectral shape with nanosheet size are observed, as indicated by the arrows in Fig. 4C. The knowledge over material size and thickness from the statistical AFM evaluation of each sample allows to quantify the changes of the extinction spectra. As discussed in more detail in previous reports on other LPE materials [15], [21], [27], [44], [45] both, the nanosheet lateral dimensions, and layer number influence the spectral shape in a specific way. For example, in transition metal dichalcogenides, peak shifts of excitonic transitions are related to differences in nanosheet thickness [21], [68], while changes in intensity ratios reflect variations in lateral size [56]. For CrTe$_3$, the analysis is limited by the absence of excitonic transitions in the spectra. Nonetheless, it is expected that intensity ratios for different wavelengths change in a systematic way with nanosheet dimensions, due to the contributions from the material edges that are electronically different from the basal planes and thus have an absorbance coefficient distinct from the basal plane. At each wavelength, a combination of both is observed which leads to size-dependent changes in the intensity profile. This behaviour can be described by a simple model assuming different extinction coefficients for the material edges and planes (equation (1) [56],

$$R = \frac{Ext_{(\lambda 1)}}{Ext_{(\lambda 2)}} = \frac{\varepsilon^{center}_{(\lambda 1)}<L>+2x(k+1)\Delta\varepsilon_{(\lambda 1)}}{\varepsilon^{center}_{(\lambda 2)}<L>+2x(k+1)\Delta\varepsilon_{(\lambda 2)}} \quad (1)$$

where is the thickness of the edge region and the nanosheet aspect ratio (length/width). A well-defined scaling of the extinction intensity ratio at 400/1000 nm with the material lateral size is observed. Equation (1) is fit to the experimental data, which describes the observed trends reasonably well (Fig. 4D). Note that similar scaling is observed for other peak intensity ratios, albeit quantitatively different. This behaviour agrees with the presented model for different electronic contributions for nanosheet edges and planes changing with sheet size and allows to derive a metric for the material's lateral size (equation (2):

$$\langle L \rangle_{Ext} = \frac{11.65 - R}{0.0173 \hat{A} \cdot R + 0.0121} \quad (2)$$

(With representing the ratio of the extinction values at wavelengths 400 and 1000 nm $R = Ext_{400nm} \hat{A} \cdot Ext_{1000nm}^{-1}$). On the one hand, this is of great practical use since it allows to assess the average lateral size of the nanosheets in unknown samples from extinction spectra. With knowledge of the scaling of lateral size and thickness (see above), the layer number can be indirectly inferred. On the other hand, it confirms that such changes in the spectral profile with nanosheet dimensions might be universal over a broad range of 2D-nanosheets.

*2.4. Nanomaterial stability*

Since our preliminary exfoliation experiments in aqueous surfactant solution pointed towards a degradation in the presence of water and oxygen, it is important to address the stability of the exfoliated nanosheets in more detail from a fundamental point of view. To test whether the structural integrity of the bulk material is retained for the nanosheets, Raman spectroscopy was applied on drop-cast nanosheets of different sizes using different excitation wavelengths. Three different sizes of nanosheets (0.4–1k $g$, 1–5k $g$ and 5–10k $g$) are shown in comparison to measurements on a bulk reference (Figure S6). The use of different excitation energies allows to identify resonance effects on the observed Raman modes and on potential decomposition products. Since the extinction spectra generally show light absorbance in the entire accessible wavelength range for all sizes of the material (see Fig. 4C), three different lasers covering a broad range of the visible spectrum (*i.e.* 457, 532 and 633 nm) were used for the measurements on both bulk- and nanomaterial. While changes of the relative intensity can be attributed to variations in the average nanosheet orientation as discussed above, no other than the expected modes are observed in all spectra. This implies that the initial structure of the material remains unchanged upon exfoliation and that insignificant oxidation occurs for the exfoliation in inert gas conditions.

To further investigate the nanosheet stability, dispersions of size-selected nanosheets were freshly prepared in an inert atmosphere and extinction spectra were acquired over time for samples stored in ambient conditions under light exclusion (Figure S7). All size-selected fractions show similar changes within the first 24 h: a signal of water arises at ~ 1430 nm, the intensity in the UV-region increases and the overall optical density decreases with respect to the initial spectrum. In addition, the formation of additional features, between 250 and 280 nm is observed. The change of the spectra seems to be qualitatively size-invariant. Hence, a polydisperse dispersion was prepared for a systematic analysis. To remove bulk-like material and impurities (or solvent degradation products that can form during sonication in pyrrolidone-based solvents) [69], a two-step centrifugation sequence was applied. Unexfoliated material was removed from the dispersion by sedimentation at 100 $g$. The supernatant, containing the material of interest was then sedimented at 30k *g*, which enables to increase the material concentration upon re-dispersion in fresh solvent and removes impurities in the supernatant. A dispersion was freshly diluted in the glovebox and stored at 20 °C in ambient conditions, excluding light for subsequent extinction measurements of the initial sample and after 24, 44, 98 and 118 h (Fig. 5 A). The trend described above for the size-selected fractions is observed for the polydisperse sample and progresses over time. An additional peak is formed at 268 nm and a dip develops at 350 nm (Fig. 5 A, inset). In addition, a water related signal intensifies over time at ~ 1450 nm (Fig. 5 B). The changes in the UV-region may be attributed to the formation of non-stoichiometric Te and Cr oxides and the increasing water concentration can be ascribed to the hygroscopicity of the dried solvent. To shine further light on the nature of the observed changes in the extinction spectra, EDX measurements on deposited nanomaterial were carried out after controlled exposure to ambient atmosphere under light exclusion (Figure S7 F). The results suggests that the initial sample exhibits a Cr:Te:O ratio of 1:3.1:0.6, which changes to 1:3.2:0.9 upon exposure to the ambient environment for 1 h. The ratio undergoes further changes to 1:2.8:8.24 after an additional 23 h, indicating a substantial increase of oxygen content in the nanomaterial upon exposure to an ambient environment.

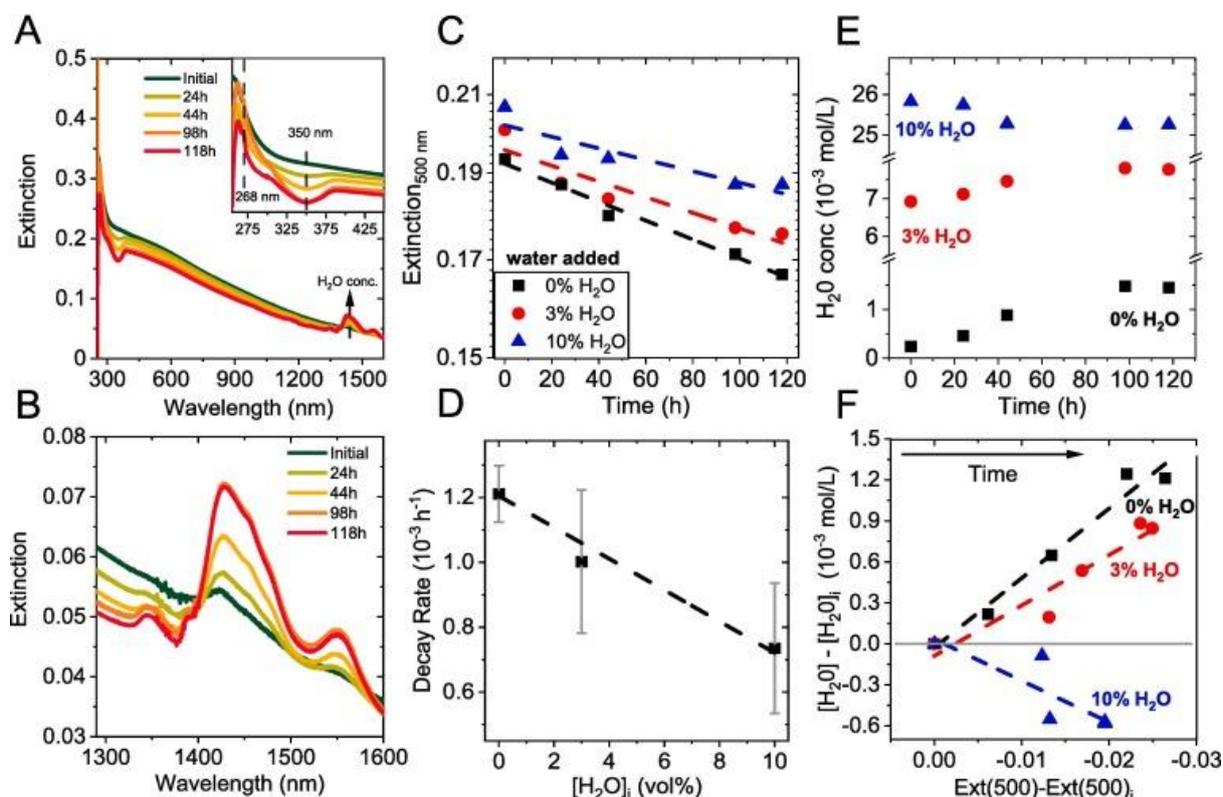

**Fig. 5.** Tracking CrTe$_3$ degradation by extinction spectroscopy. **(A)** Extinction spectra of a polydisperse CrTe$_3$ nanosheet stock dispersion stored at 20 °C in ambient conditions as function of time. The evolution of an additional feature at 268 nm and the formation of a dip at 350 nm is observed (inset). **(B)** Zoom in to the spectral region between 1300 and 1600 nm of the spectra in **(A)** showing a feature attributed to water absorption illustrating and increase

of the H₂O concentration in the solvent (dried/degassed CHP). **(C)** Extinction intensity at 500 nm (on ln scale) as function of time for three sample sets with different amounts of water added after the exfoliation (0, 3 and 10 vol%). The dashed lines are fits to (pseudo) first order decay kinetics. **(D)** Absolute extinction decay rate extracted from the slopes in (**C**) as function of the initial water concentration. The decay is slower when more water is present. **(E)** Change of the water concentration (extracted from the extinction at 1430 nm) for the three sample sets. The samples with an initial water concentration of 0 and 3 vol% show a systematic increase of the water content, while the sample with initially 10 vol% water shows a systematic decrease indicating the consumption of water on nanosheet decomposition. **(F)** Change of the water concentration versus change of the extinction at 500 nm. Only for the 10 vol%, both water and nanosheet concentration drop. For initially low water contents, more water is accumulated in the hygroscopic solvent than is consumed in the reaction with $CrTe_3$ resulting in a positive slope.

It is thus important to investigate the role of water in the reaction in more detail. To achieve this, samples containing defined initial concentrations of water (0, 3 and 10 vol%) were prepared in an identical way (nitrogen atmosphere, light exclusion) and extinction spectra were acquired over time at similar conditions. The spectra are shown in Figure S8. To analyse the spectra, the intensity at different spectral positions, *i.e.*, different wavelength was extracted to find the most suitable position for a quantification of the degradation (see Figure S9). The extinction at 500 nm, where only contributions from the initial $CrTe_3$ are observed (*i.e.*, no dips or peaks evolving over time), shows the most systematic change of the spectral profile over time. The extinction at 500 nm is plotted as a function of time in Fig. 5C (same data as in S9B–D) for the three types of samples with 0, 3 and 10 vol% of water initially added. The optical density decreases as function of time in all cases. A linearisation of the data is observed on a semi logarithmic scale suggesting (pseudo) first order kinetics. Fitting allows to estimate decay rates which were found as $1.2 \cdot 10^{-2}$, $1.0 \cdot 10^{-3}$, and $7.4 \cdot 10^{-4}$ h$^{-1}$ for the samples with 0, 3 and 10 vol% water added, respectively. The fits are reasonable in all cases, albeit with some scatter (Fig. 5C). Interestingly, the $CrTe_3$ nanosheets decay more quickly in the sample without water initially added *i.e.*, with largest rate constants. To illustrate this more clearly, the decay rates extracted as slopes from the linear fits are plotted as function of the initially added water concentration (Fig. 5D). A roughly linear decrease in the decay rate with increasing volume fraction of water initially present is observed. This is counterintuitive, as it would be expected that the rate constants increase with increasing amount of water, if water participated in the reaction as has previously been observed for liquid-exfoliated black phosphorus [43].

To investigate this degradation behaviour in more detail, the peak related to water in the extinction spectra is analysed. The effective concentration of water is estimated, using the published extinction coefficient of water in CHP at 1430 nm [43] and the background of

CrTe$_3$ at this wavelength is subtracted. Fig. 5E shows a plot of the water concentration as a function of time. The water concentration increases within the first 100 h before saturating for the samples where no water was initially added. This also applies to the sample where 3 vol% water was added prior to the degradation study. In contrast, the water concentration decreases as a function of time within the first 100 h for the sample where initially 10 vol% of water was added. This is consistent with the 10 vol% CHP-H$_2$O mixture being less hygroscopic, so that the consumption of water upon reaction with CrTe$_3$ can be spectroscopically observed. This clearly demonstrates that water participates in the reaction. Another way to represent this data is to plot the change in the water concentration as function of the change in the CrTe$_3$ extinction intensity (Fig. 5F). As the concentration of CrTe$_3$ decreases, the water content increases in the case of the 0 vol$_\%$ and 3 vol% sample but decreases for the 10 vol% sample.

The decreasing rate constant of the CrTe$_3$ degradation with increasing water content illustrated in Fig. 5D is puzzling. A possible rationale is the formation of a passivationlayer that forms relatively quickly in the presence of excess water which then slows down further degradation which is consistent with previous reports on the oxidative passivation of antimony [70]. It should also be noted that the degradation is slower and less complete than for other instable materials investigated such as black phosphorus [43], TiS$_2$ [45], MoO$_2$ [44], Ni$_2$P$_2$S$_6$ [15], and TiC [57]. This is further supported by AFM relocalisation of deposited nanosheets from a stock-like dispersion measured directly after preparation and after storage in ambient conditions for 2 days (Figure S10). To test whether the sample changes over time, the same spot on a substrate was relocated and remeasured to identify changes of the nanosheet morphology. The images show no significant change of the nanosheets which would agree with a passivation layer that quickly forms after exposure to ambient and reduces further oxidation. However, further experiments will be required to fully confirm this.

### 3. Conclusion

In summary, we demonstrate liquid exfoliation of oxidation-sensitive CrTe$_3$ under inert conditions for fabrication of a reference sample with minimal exposure to water and oxygen. Fractions of different nanosheet sizes and thicknesses were separated from an initially polydisperse distribution of nanosheet dimensions using previously established size selection protocols and quantified by statistical AFM measurements. A decrease of the nanosheet size and thickness is observed with increasing centrifugal acceleration, following a power law dependence, in line with previous reports.

The size-dependent optical response is studied by Raman and extinction measurements. Raman measurements show spot to spot variations of the intensity ratios for different vibrational modes, which we attribute to alignment effects in particular for samples deposited in drilled holes. This is confirmed by (tilt) angle dependent measurements on a crystalline bulk reference. Furthermore, a blueshift of the vibrational modes is observed after exfoliation, in reasonable agreement with DFT calculations which predict changes in Raman-active vibrations in monolayers versus bulk $CrTe_3$. In addition, photospectroscopic measurements on different fractions of the size-selected nanomaterial show a systematic change of extinction ratios with decreasing nanosheet lateral size, in agreement with a previously reported model. This enables to use the extinction ratio as a proxy for the average $CrTe_3$ nanosheet size in dispersion.

In further analysis, new features are observed in the extinction spectra after exposure of the nanosheet dispersion to ambient conditions, which is indicative for nanosheet decomposition. Titration experiments with water on the time dependent photospectroscopic response of a fresh stock dispersion reveal that water is involved during processes attributed to the material decomposition. Time dependent trends indicate a (pseudo) first order rate law for the decomposition of $CrTe_3$ nanosheets and reveal that samples with a higher water concentration show a smaller decay rate for the material decomposition, which suggests the formation of a passivation layer. This is supported further by AFM experiments for freshly deposited and relocalisation of previously measured nanosheets after 2 days in ambience. Importantly, the detailed methodology (*e.g.*, water consumption from UV–Vis spectroscopy) to investigate the degradation has not been described in literature before and can be readily transferred to other material systems that are prone to degradation. Since extinction spectrometers are widely available, we believe that this is an important demonstration for the community.

## 4. Experimental methods

*4.1. Material synthesis*

$CrTe_3$ was synthesised *via* solid state diffusion processes as reported before [22]. Stoichiometric mixtures of elemental Cr and Te were heated in evacuated and sealed quartz ampoules. The samples were heated to 573 K and were held at this temperature for 1 day. Afterwards, the temperature was increased to 723 K and the samples were annealed for 4 days. The reaction products were slowly cooled to room temperatureyielding phase-pure polycrystalline $CrTe_3$.

*4.2. Liquid phase exfoliation*

All steps were performed in an argon atmosphere, using standard Schlenk techniques. CrTe$_3$ powder was immersed in CHP (concentration c = 2 g/L, 50 mL) and sonicated over 7 h in a 100 mL round bottom flask, using a Branson ultrasonic bath (CPX2800-E, 130 W) for exfoliation. The dispersion was kept at ~ 5 °C by exchanging ice-cooled water in the bath after every 30 min to minimise any potential impact from temperature alterations. Samples prepared according to this protocol are referred to as stock dispersion.

*4.3. Size selection*

Size selection of the stock dispersion was performed using liquid cascade centrifugation. Consecutive centrifugation steps of subsequentially increasing centrifugation speeds are used in this approach, as originally reported in ref [27]. In every step, the centrifugation was performed at 20 °C for 2 h, using a Hettich Mikro 220R centrifuge equipped with a fixed-angle rotor (1195A). The respective size selection steps were 100, 400, 1000, 5000, 10,000 and 30,000 *g*, with the expression "*g*" representing the relative centrifugal force (RCF) as multiple of the earth's gravitational force. After each step, the supernatant was separated from the sediment, which was used for the next centrifugation step. The sediment was redispersed in reduced solvent volume for analysis. The expression "0.1–0.4k *g*" describes consecutive centrifugation steps. In this example, the supernatant obtained after centrifugation at 100 *g* was used for material sedimentation at 400 *g*. The expression central RCF, used in Fig. 2G-H), Figure S2D), F) and Figure S4A-C) results from the midpoint of two consecutive centrifugation steps (for 0.1–0.4k *g*, central RCF = 0.25k *g*).

*4.4. Gravimetric filtration*

The concentration of nanomaterial in each fraction was determined by filtering a known volume of dispersion onto AlOx membranes (pore size 20 nm) and washing with 1 L of dry and degassed isopropanol. Prior to weighing, membranes were dried overnight in vacuum at 60 °C before and after deposition.

*4.5. Extinction measurements*

Optical extinction spectra were acquired with an Agilent Cary 6000i spectrometer in quartz cuvettes, using 0.5 nm increments and 0.1 s integration time. The dispersions were diluted to an optical density of 0.3–0.4 at the peak.

*4.6. Raman spectroscopy*

Raman spectra were acquired with a micro-Raman spectrometer (Horiba LabRAM 800) in backscattering geometry in ambient conditions using a 457, 532 and 633 nm laser for excitation. The Raman emission was collected by a 100 × magnification long-working distance objective and dispersed by 1800 l/mm grating. The laser power was kept below 1.1 mW at all times. Typical integration times were 30 s, using 5 accumulations for each spot. Neon lines were used for calibration of the wavelength. Liquid dispersions were dropped (~5 µL) into conical indentation (Ø = 1 mm, ↓ = 0.5 mm) on an aluminiumsupport and left to dry in a nitrogen atmosphere at 60 °C overnight. The samples were protected from ambience by encapsulation with a glass slide, which was glued on top of the holes.

The tilt-angle dependent Raman measurements were performed with a Renishaw InVia-Reflex confocal Raman microscope with a 532 nm excitation laser in air under ambient conditions using a custom-made stage for tilting the samples. A $CrTe_3$ crystal was placed in the centre of rotation for the measurements. The Raman emission was collected by a 100 × magnification objective lens dispersed by a 2400 l/mm grating with 0.1% of the laser power (<1 µW). 200 spectra were accumulated with an integration time of 1 s for each measurement. The spectrometer was calibrated to a silicon reference sample prior to the measurements.

*4.7. SEM/EDX characterisation*

SEM images were acquired with a JEOL JSM-7610F field emission scanning electron microscope (FE-SEM), using an In-lens Schottky field emission electron gun with 15 kV acceleration voltage at $2.5 \cdot 10^{-9}$ mbar in a working distance of ~ 5 mm. The images were measured with a dual (upper and lower) detector system consisting of collector-, scintillator-, light guide- and photomultiplier units for secondary electron imaging (SEI).

*4.8. AFM measurements*

For atomic force microscopy, a Dimension ICON3 scanning probe microscope (Bruker AXS S.A.S.) was used in ScanAsyst mode (non-contact) in air under ambient conditions using aluminium coated silicon cantilevers (OLTESPA-R3). The concentrated dispersions were diluted with isopropanol to optical densities < 0.1 at 300 nm. A drop of the dilute dispersions (15 µL) was flash-evaporated on pre-heated (175 °C) $Si/SiO_2$ wafers (0.5x0.5 $cm^2$) with an oxide layer of 300 nm. After deposition, the wafers were rinsed with ~ 50 mL of water and ~ 50 mL of isopropanol and dried with compressed nitrogen. Typical image sizes ranged from 20x20 $\mu m^2$ for larger nanosheets to 5x5 $\mu m^2$ for small nanosheets at scan rates of 0.5–0.8 Hz with 1024 lines per image. Step height analysis was used to convert the apparent AFM thickness to layer number as discussed in the main manuscript. Previously published length corrections were used to correct lateral dimensions from cantilever broadening [38].

*4.9. XPS measurements*

For the XPS measurements the CrTe$_3$ powder as well as the exfoliated CrTe$_3$ flakes on ITOsubstrates were prepared in a N$_2$ glovebox and transferred with minimal air exposure (<30 s) to an ultrahigh vacuum chamber (ESCALAB 250Xi by Thermo Scientific, base pressure: 2·10$^{-11}$ mbar). XPS measurements were performed using an XR6 monochromated Al Kα source (hν = 1486.6 eV) and a pass energy of 20 eV. The measurements on the powder sample were performed using a flood gun in order to prevent sample charging. After the measurements of the as-prepared samples were completed, the samples were exposed to ambient air for 24 h and then re-measured using the same experimental conditions as described above.

*4.10. XRD characterisation*

Powder diffraction experiments on the bulk material were performed at photon energies of 29.2 keV (λ = 0.4246 Å) on poly-crystalline samples enclosed in sealed quartz glass capillaries at the I15 beamline at the Diamond Light Source (DLS, Oxfordshire).

*4.11. Theoretical calculations*

A density functional theory (DFT) approach as implemented in Quantum Espresso was used for all calculations [71]. For the exchange–correlation a local density approximation(LDA) was applied. Norm-conserving pseudopotentials which were taken from the old FHI PP table on the Quantum Espresso website are used. 60 Ry (816.34 eV) were used for the energy cut off. A Monkhorst-Pack grid with 8x8x1 and 6x6x6 *k* points was used for monolayer and bulk calculation, respectively. The structural optimization was performed until the forces on every atom were<0.005 eV/Å. To ensure that interlayer interactions for the monolayer structure are negligible, a unit cell with a size of 25 Å in the direction perpendicular to the stacking direction was used. A density functional perturbation theory (DFPT) approach was used for the phonon calculations.

5. **Author statement**

C.B. and K.S. designed and evaluated experiments and wrote the manuscript. K.S. and G.L.O. prepared the nanomaterial, performed AFM measurements including statistical analysis, measured SEM/EDX, UV–Vis extinction and tilt-angle dependent Raman. N.M.B., H.T. and J.M. performed Raman measurements (except for the dependence on the tilt-angle of the sample surface) and evaluated the data. S.W. performed the numerical simulations. A.H. and W.B. synthesised the starting material and measured XRD. A.W. and Y.V. measured and evaluated XPS spectra. All authors contributed to the manuscript writing and discussed the results.

**Acknowledgements**

This work was partly funded by the Deutsche Forschungsgemeinschaft (DFG, German Research Foundation) – project numbers 447264071 (INST 90/1183-1 FUGG) and 182849149 (SFB 953, B13). K.S. acknowledges financial support by DFG through SY 217/1-1. Y.V. has received funding from the European Research Council (ERC) under the European Union's Horizon 2020 research and innovation program (ERC grant agreement no. 714067, ENERGYMAPS). Computational resources used for the calculations were provided by the HPC of the Regional Computer Centre Erlangen (RRZE).We thank Jana Zaumseil for access to the infrastructure at the Chair of Applied Physical Chemistry, Heidelberg University.